\documentclass[journal]{IEEEtran}
\usepackage[latin1]{inputenc}
\usepackage{times,amsmath}
\usepackage{amssymb}
\usepackage{steinmetz}
\usepackage{pstool}
\usepackage{subfigure}
\usepackage{multirow}
\usepackage{enumerate}
\usepackage{graphicx}
\usepackage{subfig}
\usepackage{mwe}
\usepackage{bigints}
\usepackage{MnSymbol}
\usepackage{stfloats} 
\usepackage[table]{xcolor}
\usepackage[square, comma, sort&compress, numbers]{natbib}
\usepackage{algorithm,algorithmic}
\usepackage{epstopdf}
\usepackage{mathtools, cuted}

\usepackage{array}
\usepackage{color}
\usepackage{epsf}
\usepackage{epsfig}
\usepackage{amsxtra}
\usepackage{bbm}
\usepackage{authblk}
\usepackage{url}
\usepackage{adjustbox}

\usepackage{framed}
\usepackage{caption}

\newtheorem{theorem}{Theorem}
\newtheorem{lemma}{Lemma}
\newtheorem{corollary}{Corollary}

\newtheorem{remark}{Remark}
\graphicspath{ {Figures/} }

\begin{document}

\title{On the Downlink Coverage Performance of RIS-Assisted THz Networks}

\author{
Waqas Aman,~\IEEEmembership{Member,~IEEE,}, Nour Kouzayha,~\IEEEmembership{Member,~IEEE,}, Muhammad Mahboob Ur Rahman,~\IEEEmembership{Member,~IEEE,}, and Tareq Y. Al-Naffouri,~\IEEEmembership{Senior Member,~IEEE} \thanks{All the authors are with Electrical and Computer Engineering Division, King Abdullah University of Science and Technology (KAUST), Thuwal, Saudi Arabia. Emails: \{waqas.aman, nour.kouzayha, muhammad.rahman, tareq.alnaffouri\}@kaust.edu.sa.}
}


\maketitle

\begin{abstract} 
This letter provides a stochastic geometry (SG)-based coverage probability (CP) analysis of an indoor terahertz (THz) downlink assisted by a single reconfigurable intelligent surface (RIS) panel. Specifically, multiple access points (AP) deployed on the ceiling of a hall (each equipped with multiple antennas) need to serve multiple user equipment (UE) nodes. Due to presence of blockages, a typical UE may either get served via a direct link, the RIS, or both links (the composite link). The locations of the APs and blockages are modelled as a Poisson point process (PPP) and SG framework is utilized to compute the CP, at a reference UE for all the three scenarios. Monte-Carlo simulation results validate our theoretical analysis. 
\end{abstract}
\vspace{-0.5cm}
\section{Introduction}
\label{sec:intro}

Reconfigurable intelligent surfaces (RIS) which turn the hostile wireless propagation from foe to a friend, and the previously unused but enormous terahertz (THz) band (from 0.3-10 THz) are now appreciated as two key enablers for upcoming 6G cellular standard \cite{Zhang:VTM:2019}. 
Stochastic geometry (SG), on the other hand, is a theoretical tool that provides the average coverage performance of a communication system with a random/ad-hoc node geometry to gain useful system-level insights (e.g., impact of various important system parameters on the overall coverage probability) \cite{haenggi2012stochastic}. SG has been utilized to study the performance of a wide range of different wireless network configurations \cite{9378781}.

{\bf Related work.}
There exist quite a few works that do SG-based coverage probability (CP) analysis of RIS-assisted wireless communication networks. 
Authors in \cite{Tabassum:TWC:2022} compute the CP of a RIS-assisted downlink where a fraction of the user equipment (UE) nodes is considered direct users, while the remaining UEs are served by the RIS. \cite{wang2022performance} computes the CP of a RIS-assisted system where the locations of both the access points (AP) and RIS panels are modeled as a Gauss point process in the 2D space. 
\cite{Nemati:Access:2020} utilizes the SIR to compute the CP for a RIS-assisted millimeter (mm)-wave communication system. Authors in \cite{shi2022coverage} consider multiple RIS panels at random locations to compute the CP in a mm-wave cellular network.  Finally, the authors in \cite{li2022ris} compute the CP to quantify the benefit of RIS panels in the presence of random (humans and buildings) blockages, in a mm-wave cellular network. 
\textcolor{black}{To the best of authors' knowledge, the coverage performance of RIS-assisted THz downlink has not been investigated yet.}

{\bf Contributions.}
\textcolor{black}{This letter is the first study on the average coverage performance of an indoor RIS-assisted THz downlink with multiple APs, multiple UEs and a single RIS panel (amid multiple blockages). The main contributions of this work can be summarized as (i) the development of a tractable framework for studying the performance of RIS-assisted indoor THz networks (ii) deriving analytical expressions for the main performance metrics including the association probability and the CP, i.e., signal-to-interference ratio (SIR) being greater than a threshold, for the three distinct scenarios, i.e., when only the direct link is available, when only the RIS link is available, and when both links (aka composite link) are available. (iii) validating the analytical results with Monte-Carlo simulations and extracting analytical relations between the derived performance metrics and the operational parameters, including; the RIS location, and number of elements, the density and height of blockages, and the APs' density.}

\vspace{-0.3cm}
\section{System Model}
\label{sec:sys-model}
\begin{figure}[t!]
\begin{center}
	\includegraphics[width=0.78\linewidth]{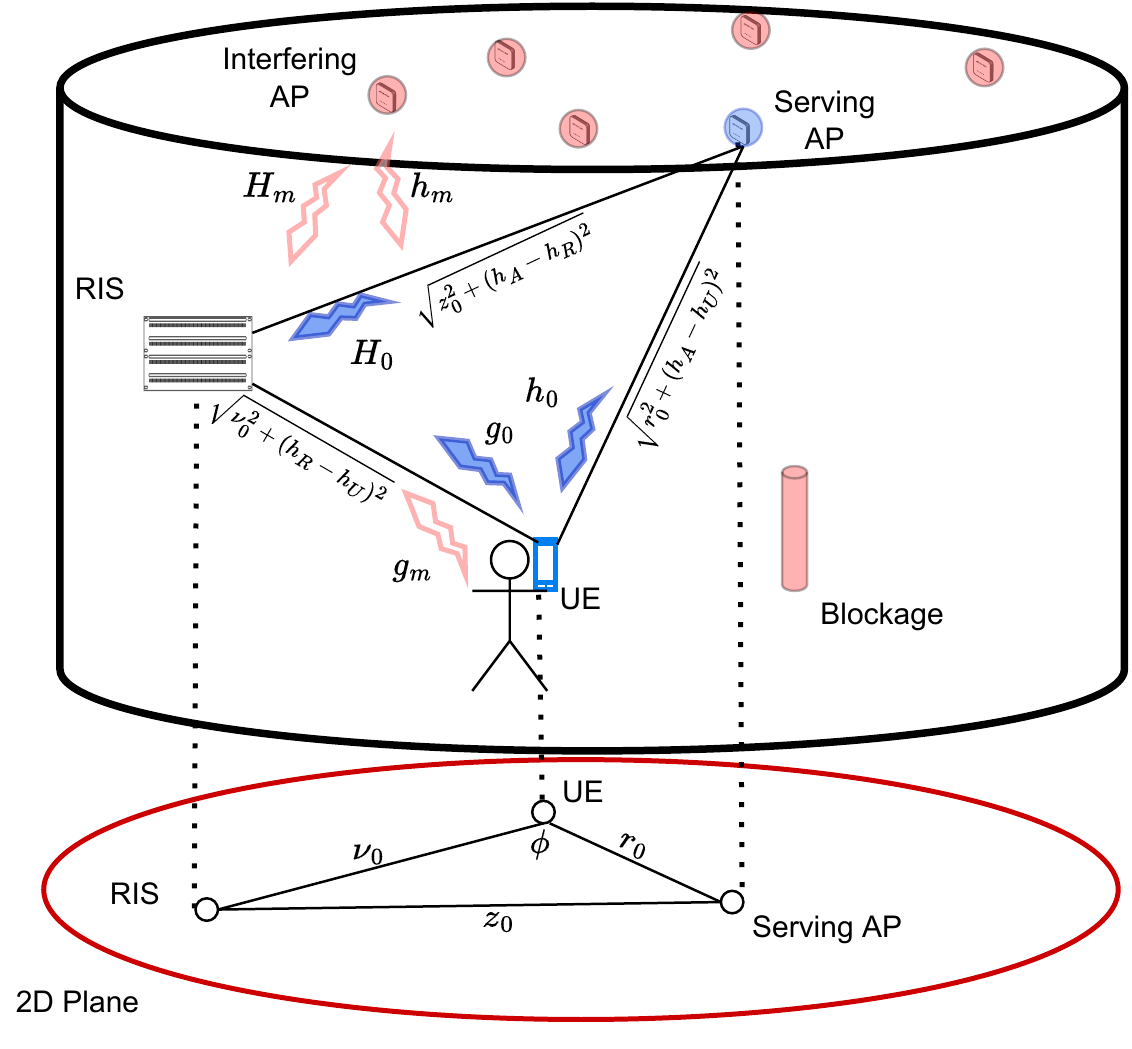} 
\vspace{-0.1cm}
\caption{System Model}
\label{fig:sysmodel}
\end{center}
\vspace{-0.9cm}
\end{figure}
\textit{Network Model:}
We consider a finite indoor THz network consisting of multiple APs deployed on the ceiling at a fixed height $h_{\mathrm{A}}$ and a single RIS panel to serve users in the downlink as shown in Fig.~\ref{fig:sysmodel}. We consider the performance of a reference user (UE) located at the origin $\bold{o}=(0,0,0)$ with a single antenna of gain $G_{\mathrm{U}}$ and at a fixed height $h_{\mathrm{U}}$ from the ground level. The locations of the APs are abstracted by a homogeneous Poisson point process (PPP) $\Pi_{\mathrm{A}}=\{x_{i}\}$ of intensity $\lambda_{\mathrm{A}}$, where $x_i$ denotes the 2D location of the $i$-th AP. The finite region $D=\bold{b}(\bold{o'},R_{t})$ is modeled as a disk of radius $R_{t}$ centered around $\bold{o'}=(0,0,h_{\mathrm{A}}-h_{\mathrm{U}})$. Each AP is equipped with $N_{\mathrm{A}}$ antennas and transmits with power $P_{\mathrm{A}}$. We assume that the RIS panel is \textcolor{black}{inside the hall and fixed} at height $h_{\mathrm{R}}$ and horizontal distance $v_{0}$ from the UE and has a total of $N=N_x \times N_y$ elements, where $N_x$ and $N_y$ are the numbers of RIS elements along the x-axis and y-axis, respectively. Finally, we assume an interference-limited scenario, and thus, ignore additive noise in our analysis and focus on the signal-to-interference-ratio (SIR) only to get the coverage probability.
\\
\textit{Blockage Model:}
We consider random blockages (humans, objects) modeled as cylinders of radius $r_{\mathrm{B}}$ and height $h_{\mathrm{B}}$ where the midpoints of cylinders are modeled as a PPP $\Pi_{\mathrm{B}}$ with intensity $\lambda_{\mathrm{B}}$. Thus, the link between the AP and the UE located at a 2D distance $r$ can be blocked and the AP-UE line-of-sight (LoS) probability is given as {\color{black}{\cite{Wu:GCW:2019}}}
\begin{equation}\small
P_\text{D}^{\text{LoS}}(r)= \exp\left(-\beta_\text{D}r\right),
\label{eq:PD_LoS}
\end{equation}
where $\beta_\text{D}=2\lambda_\text{B} r_\text{B} \vert\frac{{\color{black}\hat{h}_\text{B}}}{ {\color{black} \hat{h}_\text{A}}} \vert$, \textcolor{black}{note that the height terms with head-caps represent height terms w.r.t. $h_\text{U}=0$}.  The probability that the AP-UE link is blocked is the non-line-of-sight (NLoS) probability given as $P_\text{D}^{\text{NLoS}}(r)=1-P_\text{D}^{\text{LoS}}(r)$.
The RIS-UE link can also be blocked and the RIS-UE LoS probability is  
\vspace{-0.2cm}
\begin{equation}\small
P_\text{R}^{\text{LoS}}= \exp\left(-\beta_\text{R}v_{0}\right),
\label{eq:PR_LoS}
\end{equation}
where $\beta_\text{R}=2\lambda_\text{B} r_\text{B} \vert\frac{{\color{black}\hat{h}_\text{B}}}{{{\color{black}\hat{h}_\text{R}}}} \vert$. $P_\text{R}^{\text{NLoS}}=1-P_\text{R}^{\text{LoS}}$ is the probability that the RIS-UE link is blocked. Finally, we assume that the links between the RIS and the APs are always in a LoS condition as the RIS is higher than existing blockages. The set of non-blocked APs with direct links to the UE $\Pi_{\text{A}}$ can be obtained by thinning the PPP of APs with the LoS probability $P_\text{D}^{\text{LoS}}(r)$ and is therefore a non-homogeneous PPP with intensity $\lambda_D(r)=\lambda_A P_{\text{D}}^{\text{LoS}}(r)$.
{\color{black}\begin{remark}
If the RIS is deployed at a lower height than the existing blockages, the AP-RIS link can be blocked and the corresponding LoS probability is given as
\begin{equation}\small
P_\text{A-R}^{\text{LoS}}(z)= \exp\left(-\beta_\text{A-R}z\right),
\label{eq:PAR_LoS}
\end{equation}
where $z$ is the 2D distance separating the AP and the RIS, $\beta_\text{A-R}=2\lambda_\text{B} r_\text{B} \vert\frac{{{h}_\text{B}-{h}_\text{R}}}{{{h}_\text{A}-{h}_\text{R}}} \vert$. The probability that the AP-RIS link is blocked is the NLoS probability and is given as $P_\text{A-R}^{\text{NLoS}}(z)=1-P_\text{A-R}^{\text{LoS}}(z)$. 
\end{remark}}
\textit{Association Policy:}
The reference UE associates with the nearest AP to get service. We define a direct link as the link between the UE and the nearest AP. The RIS link denotes the indirect link to the nearest AP through the RIS panel. Due to the presence of random blockages, there are three ways a UE could get served: i) the direct link is available but the RIS link is blocked, ii) the RIS link is available but the direct link is blocked, iii) both links are available (we call such link a composite link). In all three cases, the UE receives (either direct or through-RIS) interference from the other APs.

\vspace{-0.3cm}

\section{Preliminaries: Received Signal Analysis}

We write down the expressions for the pathloss, received signal, received signal power, and interference power for the direct link, RIS link, and composite link below.

{\it 1) Direct Link:}
The pathloss for the direct link between the UE and an AP located at a 2D distance $r$ is \cite{kouzayha2021coverage}:
\begin{equation}\small
    \text{PL}_{\text{D}}(r)=\frac{G_\text{U}G_\text{A} c^2}{(4 \pi f)^2} e^{-k_a(f)\sqrt{r^2+({\color{black} \hat{h}_\text{A}})^2}} (r^2+({\color{black} \hat{h}_\text{A}})^2)^{-1
    },
    \label{eq:PL-D}
\end{equation}
where $G_\text{A}$, $G_\text{U}$ are the transmit and receive antenna's gain respectively, $f$ is the operating frequency, $c$ is the speed of light, $k_a(f)$ is the absorption coefficient, and $\sqrt{r^2+({\color{black} \hat{h}_\text{A}})^2}$ is the 3D distance between the AP and the associated UE.

When the direct line-of-sight (LoS) link is available,  the received signal at the typical UE from its serving AP at a 2D distance of $r_0$ is given by:  $y_{\text{D}}=\sqrt{P_\text{A} \text{PL}_{\text{D}}(r_0)}\mathbf{h_{0}}^T \mathbf{f_0}x$, where $P_\text{A}$ is the AP transmit power, $x$ is the transmitted symbol, $\mathbf{h_0}=[h_1,....,h_{N_{\text{A}}}]^T$ is the channel vector and $h_{j}\sim C\mathcal{N}(0,1)$ is the channel gain between the UE and the $j$-th antenna of the AP, and  $\mathbf{f_0}=[f_1,....,f_{N_{\text{A}}}]^T$ is the precoding vector.
Then, the received signal power becomes
\begin{align}
    \mathcal{S}_{\text{D}}=P_\text{A} \text{PL}_\text{D}(r_0)\vert \mathbf{h_{0}}^T \mathbf{f_0}x\vert^2.
\end{align}

The interference on the direct link is
\begin{equation}\small
       I_{\text{D}}= P_\text{A} \frac{G_\text{U}G_\text{A} c^2}{(4 \pi f)^2} \sum_{m \in \Pi_{\text{D}}/ 0} \frac{e^{-k_a(f)\sqrt{r_m^2+({\color{black} \hat{h}_\text{A}})^2}}}{(r_m^2+({\color{black} \hat{h}_\text{A}})^2)
    } \vert\mathbf{h_m}^T \mathbf{f_m}x\vert^2,
\end{equation}
where $\Pi_\text{D}$ denotes the non-homogeneous PPP (of intensity $\lambda_D$) of all the APs with direct links. 

\vspace{-0.1cm}


{\it RIS Link:}
The pathloss of the AP-RIS-UE link is given as
\begin{equation}\small
\begin{aligned}
     &\text{PL}_\text{R}(z)=
     \frac{G_\text{U}G_\text{A} (L_{x} L_{y})^2}{(4 \pi)^2(z^2+({\color{black}\hat{h}_{\text{A}}-\hat{h}_{\text{R}}})^2)(v_{0}^2+({\color{black}\hat{h}_{\text{R}}})^2)}\times \\ &e^{-k_a(f)\sqrt{z^2+({\color{black}\hat{h}_{\text{A}}-\hat{h}_{\text{R}}})^2}}e^{-k_a(f)\sqrt{v_{0}^2+({\color{black}\hat{h}_{\text{R}}})^2}}F(\theta),
\end{aligned}
\label{eq:PL-R}
\end{equation}
where $z$ is the 2D distance between the AP and the RIS, $L_x$ and $L_y$ are the length and width of a RIS element, $F(\theta)=\cos^2(\theta)=\frac{\left({\color{black}\hat{h}_{\text{A}}-\hat{h}_{\text{R}}}\right)^2}{z^2+({\color{black}\hat{h}_{\text{A}}-\hat{h}_{\text{R}}})^2}$, $\theta$ is the incident angle of the AP, $\sqrt{z^2+({\color{black}\hat{h}_{\text{A}}-\hat{h}_{\text{R}}})^2}$ is the distance between the AP and the RIS, and $\sqrt{v_{0}^2+({\color{black}\hat{h}_{\text{R}}})^2}$ is the RIS-UE distance.

The received signal through the RIS link from the serving AP located at a 2D distance $r_0$ from the UE can be expressed as $y_{\text{R}}=\sqrt{P_\text{A} \text{PL}_\text{R}\left(z_0\right)}\mathbf{g_0}^T\mathbf{\Phi}\mathbf{H_0f_0}x$, where $z_0=\sqrt{r_{0}^2+v_{0}^2-2r_{0}v_{0}\cos\phi}$ is the 2D distance separating the RIS from the serving AP, $\text{PL}_\text{R}(\cdot)$ is the total pathloss, $\mathbf{H}_0 \in \mathbb{C}^{N\times N_{\text{A}}} $ is the channel matrix between serving AP and RIS, $\mathbf{g}_0 \in \mathbb{C}^{N\times 1}$ is the RIS-UE channel vector and $\mathbf{\Phi} \in \mathbb{C}^{N\times N} $ is a diagonal matrix defined as: $\mathbf{\Phi}=\text{diag}(\text{vect}[\mathbf{\hat{\Phi}}])$ where $\text{vect}[\mathbf{A}]$ vectorizes the matrix $\mathbf{A}$, $\mathbf{\hat{\Phi}}\in \mathbb{C}^{N_x\times N_y}$ and $[\mathbf{\hat{\Phi}}]_{m,n}=e^{j\varphi_{m,n}}$, where  $\varphi_{m,n}$ is the phase shift of the $(m,n)$ RIS element~\cite{dovelos2021intelligent}.
Then, the received signal power is given as
\begin{equation}\small
    \mathcal{S}_{\text{R}}=P_\text{A} \text{PL}_\text{R}(z_0)\vert\mathbf{g_0}^T\mathbf{\Phi}\mathbf{H_0f_0}x\vert^2,
    \label{eq:SR}
\end{equation}

The interference on the RIS link is given as
\begin{equation}\small
       I_{\text{R}}= \Omega \sum_{m \in \Pi_{\text{A}} / 0} \frac{e^{-k_a(f)\sqrt{z_m^2+({\color{black}\hat{h}_{\text{A}}-\hat{h}_{\text{R}}})^2}}}{(z_m^2+({\color{black}\hat{h}_{\text{A}}-\hat{h}_{\text{R}}})^2)}  F(\theta_{m})\vert\mathbf{g_m}^T\mathbf{\Phi}\mathbf{H_mf_m}x\vert^2,
       \label{eq:IR}
\end{equation}
where $\Omega \!\!=\!\!P_\text{A} \frac{G_\text{U}G_\text{A} (L_{x} L_{y})^2 e^{-k_a(f)\sqrt{v_{0}^2+({\color{black}\hat{h}_{\text{R}}})^2}}}{(4\pi)^2 (v_{0}^2+({\color{black}\hat{h}_{\text{R}}})^2)}$, $z_m$ is the 2D distance separating the m-th AP from the RIS, and $F(\theta_m)\!=\!\cos^2(\theta_m)$, where $\theta_m$ is the incident angle for the $m$-th AP.
{\color{black}\begin{remark}
When the AP-RIS links are susceptible to blockages, not all APs can interfere with the useful transmission through RIS. The set of non-blocked APs with indirect links to the UE, denoted as $\Pi_{\text{R}}$, can be obtained by thinning the PPP of APs with the AP-RIS LoS probability $P_\text{A-R}^{\text{LoS}}(z)$ and is, therefore, a non-homogeneous PPP with intensity $\lambda_R(z)=\lambda_A P_{\text{A-R}}^{\text{LoS}}(z)$, where $z$ is the AP-RIS 2D distance. Thus, the interference on the RIS link is given as
\begin{equation}\small
       I'_{\text{R}}= \Omega \sum_{m \in \Pi_{\text{R}} / 0} \frac{e^{-k_a(f)\sqrt{z_m^2+(\hat{h}_{\text{A}}-\hat{h}_{\text{R}})^2}}}{(z_m^2+(\hat{h}_{\text{A}}-\hat{h}_{\text{R}})^2)}  F(\theta_{m})\vert\mathbf{g_m}^T\mathbf{\Phi}\mathbf{H_mf_m}x\vert^2.
\end{equation}
\end{remark}}

{\it 3) Composite Link:}
The net received signal on the composite link is the sum of the two signals received due to the two individual but parallel links (i.e., a direct link and RIS link)
\begin{equation}\small
    y_{\text{C}} \!\!= y_{\text{D}} + y_{\text{R}} 
    =\!\sqrt{P_\text{A}} \left(\sqrt{ \text{PL}_\text{D}(r_0)}\mathbf{h_{0}}^T\!\!\!\!+\!\!\sqrt{\text{PL}_\text{R}(z_0)}\mathbf{g_0}^T\!\mathbf{\Phi}\mathbf{H_0} \right)\!\mathbf{f}_0x.
\end{equation}
Thus, the net received signal power of the composite link is: 
\begin{equation}\small
    \mathcal{S}_{\text{C}}=P_{\text{A}}\vert (\sqrt{ \text{PL}_\text{D}(r_0)}\mathbf{h_{0}}^T+\sqrt{\text{PL}_\text{R}(z_0)}\mathbf{g_0}^T\mathbf{\Phi}\mathbf{H_0}) \mathbf{f}_0x \vert^2.
    \label{eq:SC}
\end{equation}
Finally, the interference on the composite link is $I_{\text{C}}=I_{\text{D}}+    I_{\text{R}}$.

\vspace{-0.2cm}

\section{Coverage Analysis}\label{sec:analysis}

\label{sec:CA}

\subsection{User Association Probability}
The association probabilities for the three coverage scenarios are given in the following lemma whose proof follows from the definition of the association scenarios and the LoS and NLoS probabilities given in (\ref{eq:PD_LoS}) and (\ref{eq:PR_LoS}).
\begin{lemma}
Conditioned on the distance $r_0$ between the UE and the nearest AP, the association probabilities through direct, RIS and composite links, denoted as $A_{\text{D}}(r_0)$, $A_{\text{R}}(r_0)$ and $A_{\text{C}}(r_0)$, respectively, are given as
\begin{equation}\small
    A_{\text{D}}(r_0)=\exp\left(-\beta_\text{D}r_{0}\right)(1-\exp\left(-\beta_\text{R}v_0\right)).
\end{equation}
\begin{equation}\small
    A_{\text{R}}(r_0)=(1-\exp\left(-\beta_\text{D}r_{0}\right))\exp\left(-\beta_\text{R}v_0\right).
\end{equation}
\begin{equation}\small
    A_{\text{C}}(r_0)=\exp\left(-\beta_\text{D}r_{0}\right)\exp\left(-\beta_\text{R}v_0\right).
\end{equation}
\begin{IEEEproof}
The association probabilities are obtained from the LoS and NLoS probabilities given in (\ref{eq:PD_LoS}) and (\ref{eq:PR_LoS}) and from the definition of the association scenarios.
\end{IEEEproof}
\label{lemma:association}
\end{lemma}
{\color{black}\begin{corollary}
When blockages exist between the RIS and the APs ($h_{\mathrm{R}}\leq h_{\mathrm{B}}$), the association probabilities through direct, RIS and composite links, denoted as $A'_{\text{D}}(r_0, z_0)$, $A'_{\text{R}}(r_0,z_0)$, and $A'_{\text{C}}(r_0,z_0)$, respectively, are given as
\begin{equation}
    A_{\text{D}}'(r_0, z_0)=\exp\left(-\beta_\text{D}r_{0}\right)\left(1-\exp\left(-\beta_\text{R}v_0-\beta_\text{A-R}z_0\right)\right),
\end{equation}
\begin{equation}
    A_{\text{R}}'(r_0,z_0)=(1-\exp\left(-\beta_\text{D}r_{0}\right))\exp\left(-\beta_\text{R}v_0\right) \exp\left(-\beta_\text{A-R}z_0\right),
\end{equation}
\begin{equation}
    A_{\text{C}}'(r_0,z_0)=\exp\left(-\beta_\text{D}r_{0}\right)\exp\left(-\beta_\text{R}v_0\right)  \exp\left(-\beta_\text{A-R}z_0\right),
\end{equation}
where $r_0$ is the 2D distance between the UE and the nearest AP, $v_0$ is the 2D RIS-UE distance and $z_0$ is 2D the distance between the RIS and the nearest AP. 
\begin{IEEEproof}
When $h_{\mathrm{R}}\leq h_{\mathrm{B}}$, the association scenarios of the UE are defined as:
\begin{itemize}
\item direct link: The link between the UE and the nearest AP is unblocked, and either the RIS-UE or the AP-RIS links, or both of them, are blocked.
\item RIS link: The direct link between the UE and the nearest AP is blocked. However, both the RIS-UE and the AP-RIS links are unblocked.
\item composite link: All the AP-UE, RIS-UE, and AP-RIS links are unblocked and the user can be served through both the direct and RIS links.
\end{itemize}
The proof follows from the definition of the LoS and NLoS probabilities of the AP-UE, AP-RIS, and RIS-UE links given in Section~\ref{sec:sys-model}. 
\end{IEEEproof}
\end{corollary}}

\subsection{Statistics of Received Signal Power and Interference}
In this section, we provide the distribution of the received signal power and Laplace transform (LT) of interference in the three coverage scenarios.

\begin{lemma}
The desired signal power through the RIS $\mathcal{S}_{\text{R}}$ given in (\ref{eq:SR}) follows the exponential distribution with parameter $\kappa_{\mathrm{R}}(z_0)=\left(2N^2(P_\text{A}\text{PL}_\text{R}(z_0)\sum_{j=1}^{N_{\text{A}}}\vert f_j\vert^2)^2\right)^{-1}$. 
\begin{IEEEproof}
We set the design parameter $\mathbf{\Phi}$ as $\mathbf{\Phi}=\text{diag}[\frac{\mathbf{g}}{\Vert \mathbf{g} \Vert}]$\footnote{\textcolor{black}{This choice of $\mathbf{\Phi}$ exploits full knowledge of the RIS-to-UE channel $\mathbf{g}$ in order to mitigate the fading effects on this channel, and thus, allows the RIS panel to do passive beamforming to the UE on the second hop.}}. Thus, the received signal through the RIS link is: 
\begin{equation}\small
\begin{aligned}
    &\mathcal{S}_{\text{R}}=P_\text{A} \text{PL}_\text{R}(z_0)\vert \mathbf{\mathbf{1}^TH_0f_0}x\vert^2 = P_\text{A}\text{PL}_\text{R}(z_{0}) \vert \sum_{j=1}^{N_{\text{A}}} f_j \sum_{i=1}^N h_{i,j} \vert^2 \\
    &=P_\text{A} \text{PL}_\text{R}(z_{0})\vert f_1 (h_{11}+h_{21}+...h_{N1})+f_2(h_{12}+h_{22}+...h_{N2}) \\ &...+f_{N_{\text{A}}}(h_{12}+h_{22}+...h_{NN_{\text{A}}})\vert^2. \\
\end{aligned}
\label{eq:SR_proof}
\end{equation}
By rearranging the summation terms inside the absolute square as $(f_1h_{11}+f_2h_{12}....+f_{N_{\text{A}}}h_{1N_{\text{A}}})....+(f_1h_{N1}+f_2h_{N2}....+f_{N_{\text{A}}}h_{NN_{\text{A}}}) $ and letting $Z_i= (f_1h_{i1}+f_2h_{i2}....+f_{N_{\text{A}}}h_{iN_{\text{A}}})$, the dual-summation term in (\ref{eq:SR_proof}) can be written as $\sum_{j=1}^{N_{\text{AP}}} f_j \sum_{i=1}^N h_{i,j} = \sum_{i=1}^N Z_i$, where $Z_{i}\sim C\mathcal{N}(0,\sum_{j=1}^{N_{\text{A}}} \vert f_j \vert^2 )$. The summation in (\ref{eq:SR_proof}) is equivalent to a summation of $N$ i.i.d. random variables and follows the $C\mathcal{N}(0,N\sum_{j=1}^{N_{\text{A}}}\vert f_j\vert^2)$ distribution. Finally, the term $\vert \mathbf{g}^H\mathbf{\Phi}\mathbf{H_0f_0}\vert$ is Rayleigh distributed with parameter $N\sum_{j=1}^{N_{\text{A}}}\vert f_j\vert^2$ and $\mathcal{S}_{\text{R}}$ is exponentially distributed with parameter $\kappa_{\text{R}}(z_{0}) =\frac{1}{2N^2(P_\text{A}\text{PL}_\text{R}(z_{0}) \sum_{j=1}^{N_{\text{A}}}\vert f_j\vert^2)^2}$.
\end{IEEEproof}
\label{lemma:SR}
\end{lemma}

The distribution of the desired signal power through the composite link $\mathcal{S}_{\text{C}}$ is given in the following lemma whose proof follows a similar procedure as in Lemma~\ref{lemma:SR}.
\begin{lemma}
The desired signal power through the composite link  $\mathcal{S}_{\text{C}}$ given in (\ref{eq:SC}) follows the exponential distribution with parameter $\kappa_{\mathrm{C}}(r_{0},z_{0})=\frac{1}{(\sqrt{2}(N\text{PL}_\text{R}(z_{0})+\text{PL}_\text{D}(r_{0}))\sum_{j=1}^{N_{\text{A}}}\vert f_j \vert^2)^2}$.

\end{lemma}

Next, we derive the LT of interference from direct APs $I_{\mathrm{D}}$, through the RIS $I_{\mathrm{R}}$ and through composite links $I_{\mathrm{C}}$ in the following Lemmas. 
\begin{lemma}
Conditioning on the 2D distance $r_0$ between the UE and the serving AP, the LT of the net interference from all APs with direct links with respect to the reference UE is
\begin{equation}\small
\begin{aligned}
    \mathcal{L}_{{I}_{\text{D}}}(s)=
    \exp\left[-2\pi\lambda_{\mathrm{A}} \int_{r_{0}}^{R_{t}}\!\left(1-\frac{\kappa_{\text{D}}}{\kappa_{\text{D}}+s P_{\text{A}}\text{PL}_{\mathrm{D}}(r)}\right)\!P_{\mathrm{D}}^{\mathrm{LoS}}(r)r\text{d}r\right],
\end{aligned}
\label{eq:LT_I_D}
\end{equation}
where $\kappa_{\text{D}}=\frac{1}{2(\sum_{j=1}^{N_{\text{A}}}\vert f_j\vert^2)^2}$,  $P_{\mathrm{D}}^{\mathrm{LoS}}(\cdot)$ is LoS probability given in (\ref{eq:PD_LoS}), and $\text{PL}_{\mathrm{D}}(\cdot)$ is the pathloss of the direct link given in (\ref{eq:PL-D}).
\begin{IEEEproof}
The Laplace transform $\mathcal{L}_{{I}_{\text{D}}}(s)$ can be derived as
\vspace{-0.2cm}
\begin{equation}\small
\begin{aligned}
    \mathcal{L}_{{I}_{\text{D}}}(s) =\mathbb{E}_{{I}_{\text{D}}}\left[e^{-s{I}_{\text{D}}}\right] 
     &\stackrel{(a)}{=}\mathbb{E}_{\Pi_\text{D},\gamma_m}\left[\exp\left(-s\!\!\!\!\sum_{m\in \Pi_{\text{D}}/0}\!\!\! P_{\mathrm{A}}\text{PL}_{\mathrm{D}}(r_m)\gamma_m\right)\right]\\
    &\stackrel{(b)}{=}\mathbb{E}_{\Pi_\text{D}}\left[\prod_{m\in \Pi_{\text{D}/0}} \frac{\kappa_\text{D}}{\kappa_\text{D}+s P_{\text{A}}\text{PL}_{\mathrm{D}}(r_m)}\right],
    \end{aligned}
    \vspace{-0.1cm}
\end{equation}
where (a) is obtained by defining $\gamma_m=\vert \mathbf{f}_m \mathbf{h}_m \vert^2$, and (b) is obtained from the distribution of $\gamma_m$ which follows the exponential distribution with parameter $\kappa_{\text{D}}=\frac{1}{2(\sum_{j=1}^{N_{\text{A}}}\vert f_j\vert^2)^2}$. The expression in (\ref{eq:LT_I_D}) is finally obtained by omitting the index $m$ and after applying the probability generating functional (PGFL) on the PPP of direct APs with intensity $\lambda_{\text{A}}P_{\mathrm{D}}^{\mathrm{LoS}}(r)$.
 
\vspace{-0.1cm}
\end{IEEEproof}
\end{lemma}
\begin{lemma}
Conditioning on the 2D distance $r_0$ between the UE and the serving AP, the LT of the aggregate interference from all APs having non-blocked links through the RIS with respect to the reference UE is given as
\begin{equation}\small
\begin{aligned}
   &\mathcal{L}_{{I}_{\text{R}}}(s)=\exp\Bigg[-2\pi\lambda_{\mathrm{A}}\\
   &\int_{0}^{\pi}\!\!\int_{r_0}^{R_t}\!\!\left(1-Q\left(s P_{\text{A}}\text{PL}_{\mathrm{R}}\left(\sqrt{r^2+v_{0}^2-2 r v_{0} \cos\phi}\right)\right)\right)r\text{d}r\text{d}\phi\Bigg], 
\end{aligned}
  \label{eq:LT_I_RIS}
\end{equation}
where $Q(x)=\frac{1}{\sqrt{1+2x}}\exp\left(-\frac{\kappa_{\text{I}}x}{1+2x}\right)$, $\kappa_{\text{I}}=\frac{\mu_{B}}{2\left(\sum_{j=1}^{N_{\text{A}}}\vert f_j^{_m} \vert \right)\sigma_{B}^2}$, $\mu_{B}=\frac{\pi}{2}$, $\sigma_{B}^2=2^2\left(1-\frac{\pi^2}{16}\right)$, $\text{PL}_{\mathrm{R}}(\cdot)$ is the pathloss of the direct link given in (\ref{eq:PL-R}) and $v_0$ is the 2D  RIS-UE distance.
\begin{IEEEproof}
First, we derive the distribution of the term $\zeta_m=\vert\mathbf{g_m}^T\mathbf{\Phi}\mathbf{H_mf_m}\vert^2=\vert \sum_{j=1}^{N_{\text{A}}} f_j^{m} \sum_{i=1}^N g_i^mh_{i,j}^m e^{-\iota \psi_i}\vert^2$ in the interference through RIS link expression in (\ref{eq:IR}). Using the polar form of the complex numbers and by letting $B_{i,j}^m=\vert g_i^m \vert \vert h_{i,j}^m \vert$, and $\tau_{i,j}^m=\theta_{g_i^m}+\theta_{h_{i,j}^m}-\psi_i$, we can write the summation as
\vspace{-0.2cm}
\begin{equation}\small
\begin{aligned}
     &\sum_{j=1}^{N_{\text{A}}} \vert f_j^{m} \vert e^{\iota \theta_{f_{j}^m}} \sum_{i=1}^N \vert g_i^m \vert \vert h_{i,j}^m \vert e^{\iota (\theta_{g_i^m}+\theta_{h_{i,j}^m}-\psi_i)}\\
     &= \sum_{j=1}^{N_{\text{A}}} \vert f_j^{m} \vert e^{\iota \theta_{f_j}^m} \sum_{i=1}^N B_{i,j}^m e^{\iota \tau_{i,j}^m}\stackrel{(a)}{\leq}\sum_{i=1}^{N}B_{i}^m\sum_{j=1}^{N_{\text{A}}}\vert f_j^{m} \vert,
\end{aligned}
\label{eq:IR_channel}
\vspace{-0.1cm}
\end{equation}

where (a) is obtained by replacing $B_{i,j}$ with $B_i$ since $B_{i,j}^m\ \forall i,j$ are i.i.d. Note that $B_{i,j}^m$ is the product of two independent Rayleigh distributed random variables and has a mean $\mu_{B}=\frac{\pi}{2}$, and variance $\sigma_{B}^2=2^2\left(1-\frac{\pi^2}{16}\right)$. The expression in (\ref{eq:IR_channel}) can be expressed as $\sum_{i=1}^N C_{N_{\text{A}}} B_i$,
where $C_{N_{\text{A}}}=\sum_{j=1}^{N_{\text{A}}}\vert f_j^{_m} \vert $.
As $N$ is significantly large in a typical RIS panel, we can invoke the central limit theorem (CLT) and $\sum_{i=1}^N C_{N_{\text{A}}}B_i$ will follow a normal distribution with mean $NC_{N_{\text{A}}}\mu_B$ and variance $N C_{N_{\text{A}}}^2\sigma_B^2$. Consequently, $\vert\mathbf{g_m}^T\mathbf{\Phi}\mathbf{H_mf_m}x\vert^2$ follows a non-central chi-squared distribution $\chi^2_1(\kappa_{\text{I}})$ with one degree of freedom and non-centrality parameter $\kappa_{\text{I}}=\frac{1}{2}\frac{\mu_B}{C_{N_{\text{A}}}\sigma_B^2}$.

The LT of RIS interference is derived as
 \begin{equation}\small
\begin{aligned}
     &\mathcal{L}_{I_{\text{R}}}\!(s) \!=
     \mathbb{E}_{I_{\text{R}}}\!\left[\exp\left(-s\Omega \!\!\!\!\! \sum_{m \in \Pi_{\text{A}}/0}\!\!\!\!\! \frac{e^{-k_a(f)\sqrt{z_{m}^2+({\color{black}\hat{h}_{\text{A}}-\hat{h}_{\text{R}}})^2}}({\color{black} \hat{h}_\text{A}-{\hat{h}_\text{R}}})^2}{(z_{m}^2+({\color{black}\hat{h}_{\text{A}}-\hat{h}_{\text{R}}})^2)^2} \zeta_m \!\right)\right] \\
     &\stackrel{(a)}{=}\mathbb{E}_{\Pi_{\text{A}}}\left[\prod_{m \in \Pi_{\text{A}}/0}\!\!\!\!Q\left(s\Omega\frac{e^{-k_a(f)\sqrt{z_{m}^2+({\color{black}\hat{h}_{\text{A}}-\hat{h}_{\text{R}}})^2}}({\color{black} {\color{black}\hat{h}_\text{A}-\hat{h}_\text{R}}})^2}{(z_{m}^2+({\color{black}\hat{h}_{\text{A}}-\hat{h}_{\text{R}}})^2)^2}\right)\right], 
\end{aligned}
\end{equation}
where (a) is obtained from the independence of $\zeta_m=\vert\mathbf{g_m}^T\mathbf{\Phi}\mathbf{H_mf_m}\vert^2$ and $\Pi_{\mathrm{A}}$ and from the moment generating function (MGF) $Q(x)$ of $\zeta_m$ which follows the non-central Chi-square distribution, i.e., $Q(x)=\frac{1}{\sqrt{1+2x}}\exp\left(-\frac{\kappa_{\text{I}}x}{1+2x}\right)$. The final expression of $\mathcal{L}_{I_{\text{R}}}(s)$ given in (\ref{eq:LT_I_RIS}) is obtained by applying the PGFL on the APs PPP $\Pi_{\text{A}}$ and by reformulating $z_m$ using the cosine rule as a function of the angle and 2D distance from UE to $m$-th AP, i.e $z_{m}=\sqrt{r_m^2+v_{0}^2-2 r_m v_{0} \cos\phi_m}$.

\end{IEEEproof}
\label{lemma:LIR}
\end{lemma}

\begin{corollary}
The LT of interference from composite links includes the interference from direct links and RIS links and is given as $\mathcal{L}_{{I}_{\text{C}}}(s)=\mathcal{L}_{{I}_{\text{D}}}(s)\mathcal{L}_{{I}_{\text{R}}}(s)$,
where $\mathcal{L}_{{I}_{\text{D}}}(s)$ and $\mathcal{L}_{{I}_{\text{R}}}(s)$ are given in (\ref{eq:LT_I_D}) and ($\ref{eq:LT_I_RIS}$), respectively.
\end{corollary}
{\color{black}\begin{corollary}
The LT of the aggregate interference from all APs having non-blocked links through the RIS when blockages are considered on the AP-RIS links is given as
\begin{equation}\small
\begin{aligned}
   &\mathcal{L}_{{I'}_{\text{R}}}(s)=\exp\Bigg[-2\pi\lambda_{\mathrm{A}}\int_{0}^{\pi}\!\!\int_{r_0}^{R_t}\\
   &\!\!\left(1-Q\left(s P_{\text{A}}\text{PL}_{\mathrm{R}}\left(\sqrt{r^2+v_{0}^2-2 r v_{0} \cos\phi}\right)\right)\right)P_{A-R}^{\text{LoS}}(r,\phi)r\text{d}r\text{d}\phi\Bigg], 
\end{aligned}
\end{equation}
where $Q(x)$ is given in Lemma~\ref{lemma:LIR} and $P_{A-R}^{\text{LoS}}(r,\phi)$ is the LoS probability on the AP-RIS links and is given as $P_{A-R}^{\text{LoS}}(r,\phi)=\exp\left(-\beta_\text{A-R}\sqrt{r^2+v_{0}^2-2 r v_{0} \cos\phi}\right)$. 
\begin{IEEEproof}
    The proof follows the same steps of Lemma~\ref{lemma:LIR}, thus omitted.
\end{IEEEproof}
\end{corollary}}

\vspace{-0.3cm}
\subsection{Coverage Probability}
Since the UE has three different coverage scenarios, the overall coverage probability is derived using the law of total probability and is given in the following theorem.
\begin{theorem}
The coverage probability of the reference UE in a RIS-assisted downlink THz network is given as
\begin{equation}\small
\begin{aligned}
    &P_{\text{Cov}}=\int_{0}^{\pi}\frac{1}{\pi}\int_{0}^{R_t}f_{r_{0}}(r_{0})\left(A_{\text{D}}(r_{0})P_{\text{D}}(r_{0})\right.\\
    &\left.+A_{\text{R}}(r_{0})P_{\text{R}}(r_{0},\phi_{0})+A_{\text{C}}(r_{0})P_{\text{C}}(r_{0},\phi_{0})\right)\text{d}r_0\text{d}\phi_0,  
\end{aligned}
\end{equation}
where $f_{r_{0}}(r_{0})$ is the PDF of the 2D distance separating the UE from the nearest AP and is given as $f_{r_{0}}(r_0)=2\pi\lambda_{\text{A}}r_{0}\exp\left(-\lambda_{\text{A}}\pi r_{0}^2\right)$, $A_{\text{D}}(r_{0})$, $A_{\text{R}}(r_{0})$ and $A_{\text{C}}(r_{0})$ are given in Lemma~\ref{lemma:association} and $P_{\text{D}}(r_{0})$, $P_{\text{R}}(r_{0},\phi_0)$ and $P_{\text{C}}(r_{0},\phi_0)$ are the conditional coverage probabilities given the three coverage scenarios and are expressed as
\vspace{-0.3cm}
\begin{equation}\small
\label{eq:PD}
    P_{\text{D}}(r_{0})=\mathcal{L}_{{I}_{\text{D}}}\left(\frac{\tau\kappa_\text{D}}{P_\text{A} \text{PL}_\text{D}(r_0)}\right),
\end{equation}
\begin{equation}\small
    P_{\text{R}}(r_{0},\phi_0)=\mathcal{L}_{{I}_{\text{D}}}\left(\tau\kappa_\text{R}(z_{0})\right)\mathcal{L}_{{I}_{\text{R}}}(\tau\kappa_\text{R}(z_{0})),
    \label{eq:PR}
\end{equation}
\begin{equation}\small
    P_{\text{C}}(r_{0},\phi_0)=\mathcal{L}_{{I}_{\text{D}}}\left(\tau\kappa_\text{C}(r_{0},z_{0})\right)\mathcal{L}_{{I}_{\text{R}}}(\tau\kappa_\text{C}(r_{0},z_{0})),
    \label{eq:PC}
\end{equation}
where $z_{0}=\sqrt{r_{0}^2+v_{0}^2-2r_{0}v_{0}\cos\phi_{0}}$, $\mathcal{L}_{{I}_{\text{D}}}(\cdot)$ and $\mathcal{L}_{{I}_{\text{R}}}(\cdot)$ are the Laplace transform of interference through the direct and the RIS links and are given in (\ref{eq:LT_I_D}) and (\ref{eq:LT_I_RIS}), respectively.
\begin{IEEEproof}
The conditional coverage probability for a direct link $P_{\text{D}}(r_0)$ is defined as the SIR on the direct link is higher than a predefined threshold $\tau$ and can be derived as
\begin{equation}\small
    \begin{aligned}
    &P_{\text{D}}(r_0)=\mathbb{P}\left[\frac{\mathcal{S}_{\text{D}}}{I_{\text{D}}}>\tau\right]=\mathbb{P}\left[\vert \mathbf{h_{0}}^T \mathbf{f_0}\vert^2>\frac{\tau I_{\text{D}}}{P_\text{A} \text{PL}_{\text{D}}(r_{0})}\right]\\
    &\stackrel{(a)}{=}\mathbb{E}_{I_{\text{D}}}\left[\exp\left(-\kappa_\text{D} \tau\frac{ {I_{\text{D}}}}{P_\text{A} \text{PL}_\text{D}(r_{0})}\right)\right] 
     \stackrel{(b)}{=}\mathcal{L}_{{I}_{\text{D}}}\left(\frac{\tau\kappa_\text{D}}{P_\text{A} \text{PL}_\text{D}(r_0)}\right),
\end{aligned}
\end{equation}
where (a) follows from the complementary cumulative distribution function (CCDF) of the channel and antenna gain parameter $\vert \mathbf{h_{0}}^T \mathbf{f_0}\vert^2$ which follows the exponential distribution with parameter $\kappa_{\text{D}}=\frac{1}{2(\sum_{i=1}^{N_{\text{A}}}\vert f_i\vert^2)^2}$ and (b) follows from the definition of LT of interference from APs with direct links $I_{\text{D}}$.

When, the RIS link is not blocked, the UE can receive interference from all APs through the RIS and from APs with non-blocked direct links. Thus, the conditional coverage probability for a RIS link can be derived as
\begin{equation}\small
\begin{aligned}
 P_{\text{R}}(r_0,\phi_0)&=\!\mathbb{P}\left[\frac{\mathcal{S}_{\text{R}}}{I_{\text{R}}+I_{\text{D}}}\!\!>\!\tau\right]\stackrel{(a)}{=}\mathbb{E}\left[\exp{(-\tau\kappa_\text{R}(z_0)( I_{\text{R}}+I_{\text{D}}))}\right],\\
\end{aligned}
\end{equation}
where (a) is obtained from the exponential distribution of the received signal power $\mathcal{S}_{\text{R}}$ derived in Lemma~\ref{lemma:SR}. The final expression in (\ref{eq:PR}) is obtained using the definition of Laplace transform and the independence of direct and RIS interference ${I_{\text{D}}}$ and ${I_{\text{R}}}$ when conditioned on the 2D distance and angle to nearest AP $r_0$ and $\phi_{0}$. The conditional coverage probability for the composite link is obtained following the same procedure and the final expression is given in~(\ref{eq:PC}). Finally, the overall coverage probability can be obtained by using the total law of probability and deconditioning with respect to $r_{0}$ and $\phi_0$ which has a uniform distribution in $[0,\pi]$.
\end{IEEEproof}
\label{theorem:coverage}
\end{theorem}
{\color{black}\begin{corollary}
  The coverage probability in the case of blockages existence on the AP-RIS links ($h_\text{R}\leq h_\text{B}$) is given as
  \begin{equation}\small
\begin{aligned}
    P'_{\text{Cov}}&=\int_{0}^{\pi}\frac{1}{\pi}\int_{0}^{R_t}f_{r_{0}}(r_{0})\Big(A'_{\text{D1}}(r_{0},\phi_{0})P'_{\text{D1}}(r_{0})\\
    &+A'_{\text{D2}}(r_{0},\phi_{0})P'_{\text{D2}}(r_{0})+A'_{\text{R}}(r_{0},\phi_{0})P'_{\text{R}}(r_{0},\phi_{0})\\
    &+A'_{\text{C}}(r_{0},\phi_{0})P'_{\text{C}}(r_{0},\phi_{0})\Big)\text{d}r_0\text{d}\phi_0,  
\end{aligned}
\end{equation}
where $A'_{\text{D1}}(r_0,\phi_0)=\exp\left(-\beta_{\text{D}}r_0-\beta_{\text{R}}v_0\right)\left(1-\exp\left(-\beta_{\text{A-R}}z_0\right)\right)$, $A'_{\text{D2}}(r_0)=\exp\left(-\beta_{\text{D}}r_0\right)\left(1-\exp\left(-\beta_{\text{R}}v_0\right)\right)$, and $z_0=\sqrt{r_0^2+v_0^2-2r_0v_0\cos\phi_0}$.
$P'_{\text{D1}}(r_0)$, $P'_{\text{D2}}(r_0)$, $P'_{\text{R}}(r_{0},\phi_{0})$, and $P'_{\text{C}}(r_0,\phi_0)$ are the conditional coverage probabilities and are given as
\begin{equation}
    P'_{\text{D1}}(r_0)=\mathcal{L}_{{I}_{\text{D}}}\left(\frac{\tau\kappa_\text{D}}{P_\text{A}\text{PL}_\text{D}(r_0)}\right),
\end{equation} 
\begin{equation}
    P'_{\text{D2}}(r_0)=\mathcal{L}_{{I}_{\text{D}}}\left(\frac{\tau\kappa_\text{D}}{P_\text{A} \text{PL}_\text{D}(r_0)}\right) \mathcal{L}_{{I'}_{\text{R}}}\left(\frac{\tau \kappa_\text{D}}{P_\text{A} \text{PL}_\text{D}(r_0)}\right),
\end{equation} 
\begin{equation}
   P'_{\text{R}}(r_{0},\phi_{0})=\mathcal{L}_{{I}_{\text{D}}}\left(\tau\kappa_\text{R}(z_{0})\right)\mathcal{L}_{{I'}_{\text{R}}}(\tau\kappa_\text{R}(z_{0})), 
\end{equation}
\begin{equation}
P'_{\text{C}}(r_{0},\phi_{0})=\mathcal{L}_{{I}_{\text{D}}}\left(\tau\kappa_\text{C}(r_0, z_{0})\right)\mathcal{L}_{{I'}_{\text{R}}}(\tau\kappa_\text{C}(r_0, z_{0})),
\end{equation}
and $\mathcal{L}_{I_{\mathrm{D}}}(\cdot)$ and $\mathcal{L}_{{I'}_{\mathrm{R}}}(\cdot)$ are the Laplace transforms of interference from the direct and indirect through RIS APs.
\begin{IEEEproof}
While the derivations of the conditional coverage probabilities for the RIS link and the composite link scenarios remain unchanged when blockages are considered on the AP-RIS links, the conditional coverage probability for the direct link accounts for two different cases:
\begin{itemize}
    \item The AP-UE and RIS-UE links are not blocked. However, the AP-RIS link is blocked. In this case, the interference received at the UE accounts for the APs with unblocked direct links and indirect links through the RIS. Thus, the interference can be expressed as $I'=I_{\mathrm{D}}+I'_{\mathrm{R}}$. The probability of occurrence of such event is given as $A'_{\mathrm{D1}}(r_0,z_0)=P_{\mathrm{D}}^{\mathrm{LoS}}(r_0)P_{\mathrm{R}}^{\mathrm{LoS}}\left(1-P_{\mathrm{A-R}}^{\mathrm{LoS}}(z_0)\right)$.
    \item The AP-UE link is not blocked and the RIS-UE link is blocked. In this case, the interference is caused by only the APs with unblocked direct links. The interference can be expressed as $I'=I_{\mathrm{D}}$. The probability of occurrence of such event is given as $A'_{\mathrm{D2}}(r_0)=P_{\mathrm{D}}^{\text{LoS}}(r_0)\left(1-P_{\mathrm{R}}^{\mathrm{LoS}}\right)$.
\end{itemize}
The remaining proof follows a similar procedure as in Theorem~\ref{theorem:coverage}, thus omitted. 
\end{IEEEproof}
\end{corollary}}

\begin{figure}[t]
    \centering
    \includegraphics[scale=0.6]{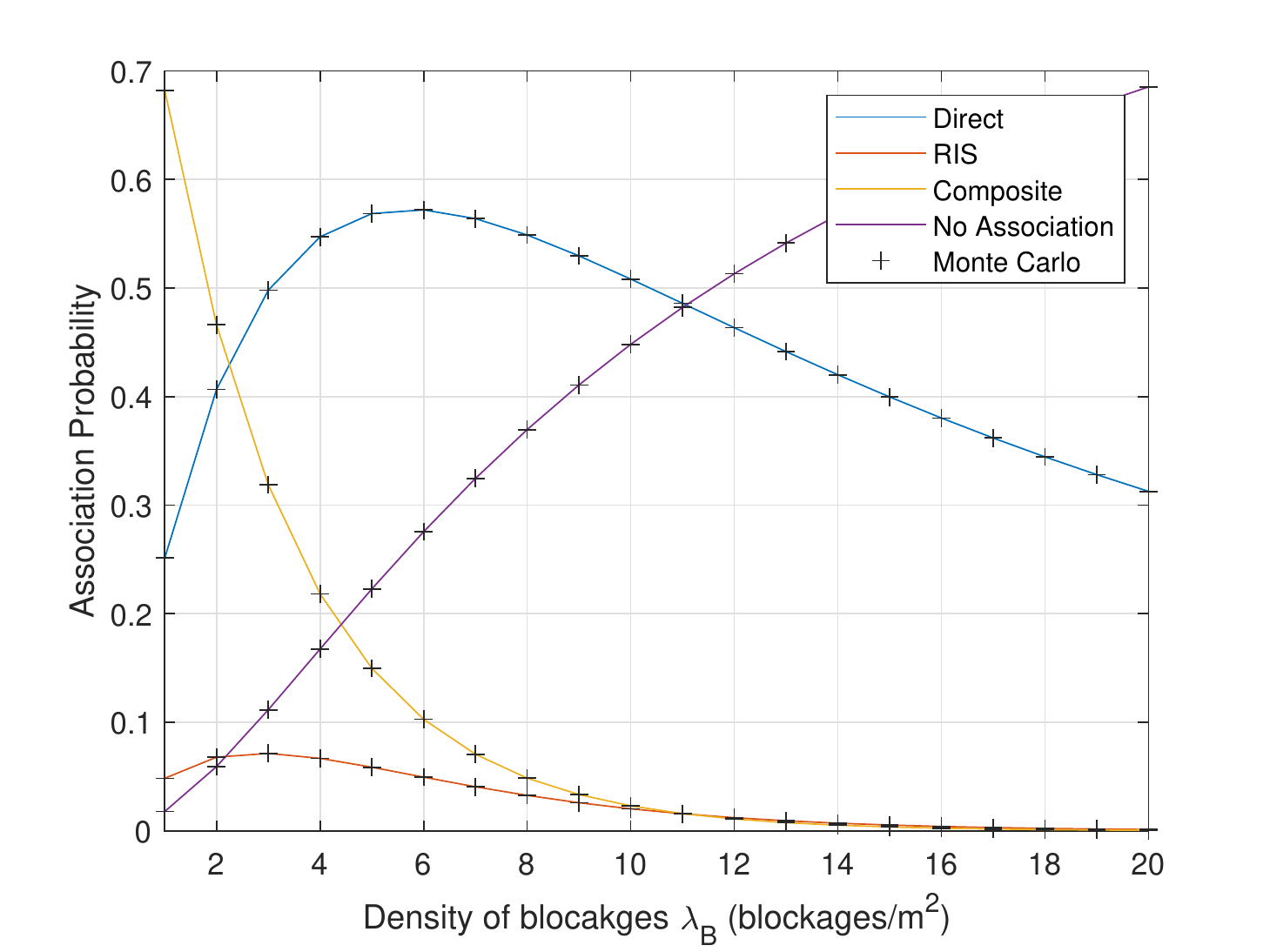}
    \caption{Association probabilities as a function of $\lambda_\text{B}$}
    \label{fig:Ap_v_LB}
\end{figure}


\begin{figure}[t]
    \centering
    \includegraphics[scale=0.6]{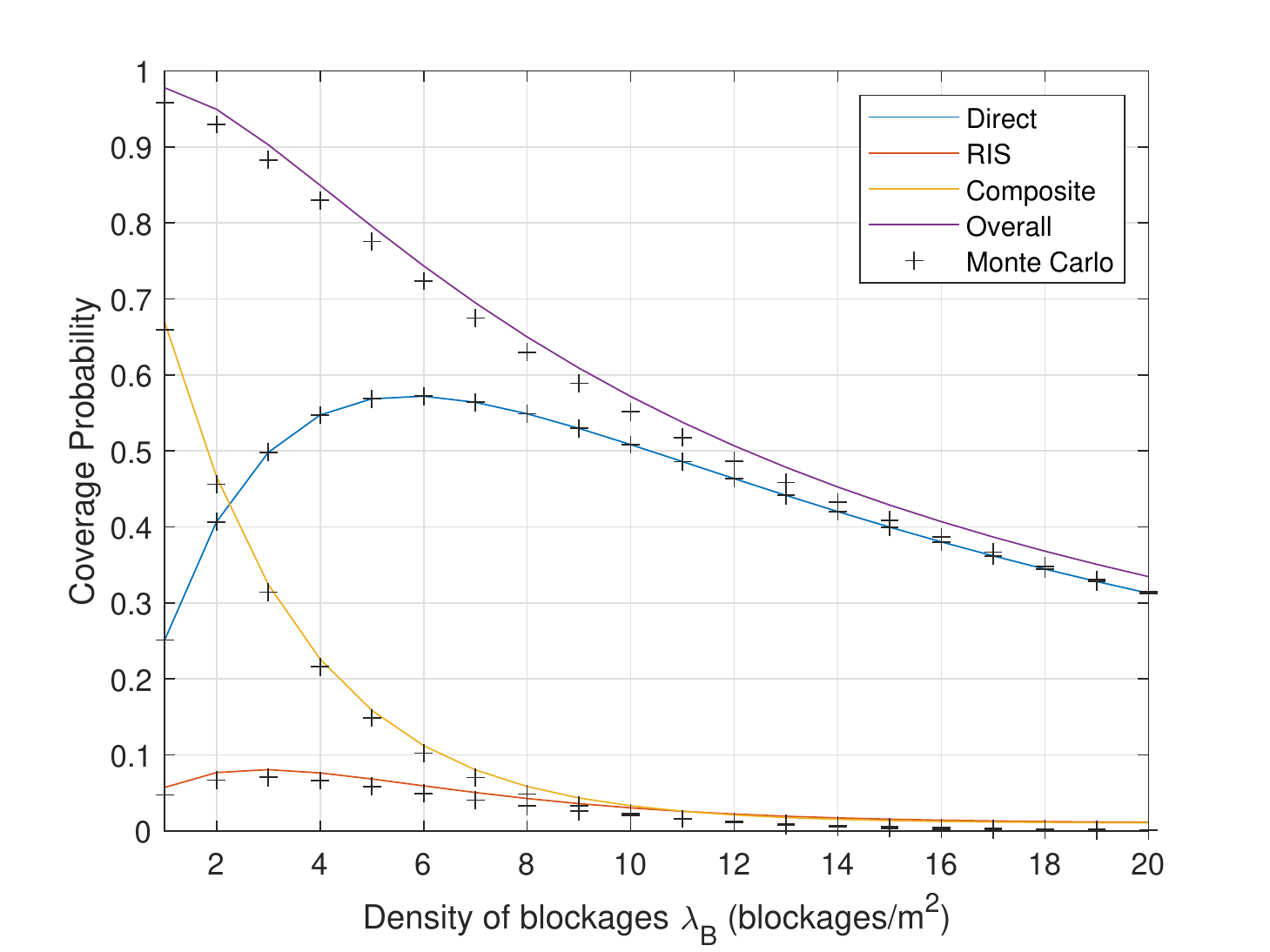}
    \caption{Coverage probabilities as a function of $\lambda_\text{B}$}
    \label{fig:CP_v_LB}
\end{figure}


\begin{figure}[t]
    \centering
    \includegraphics[scale=0.5]{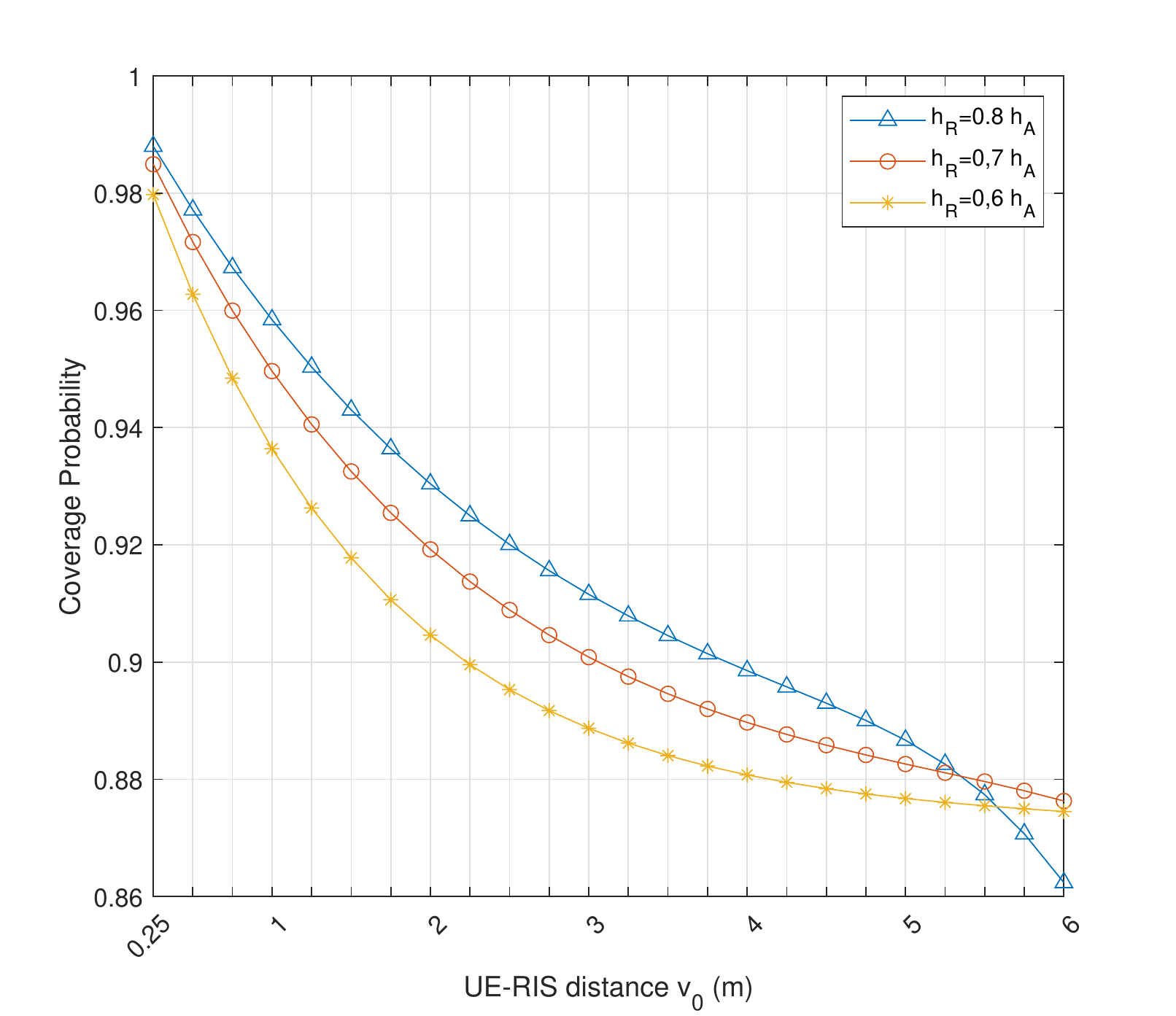}
    \caption{Coverage probability as a function of the UE-RIS distance and the RIS height.}
    \label{fig3:a}
\end{figure}


\begin{figure}[t]
    \centering
    \includegraphics[scale=0.6]{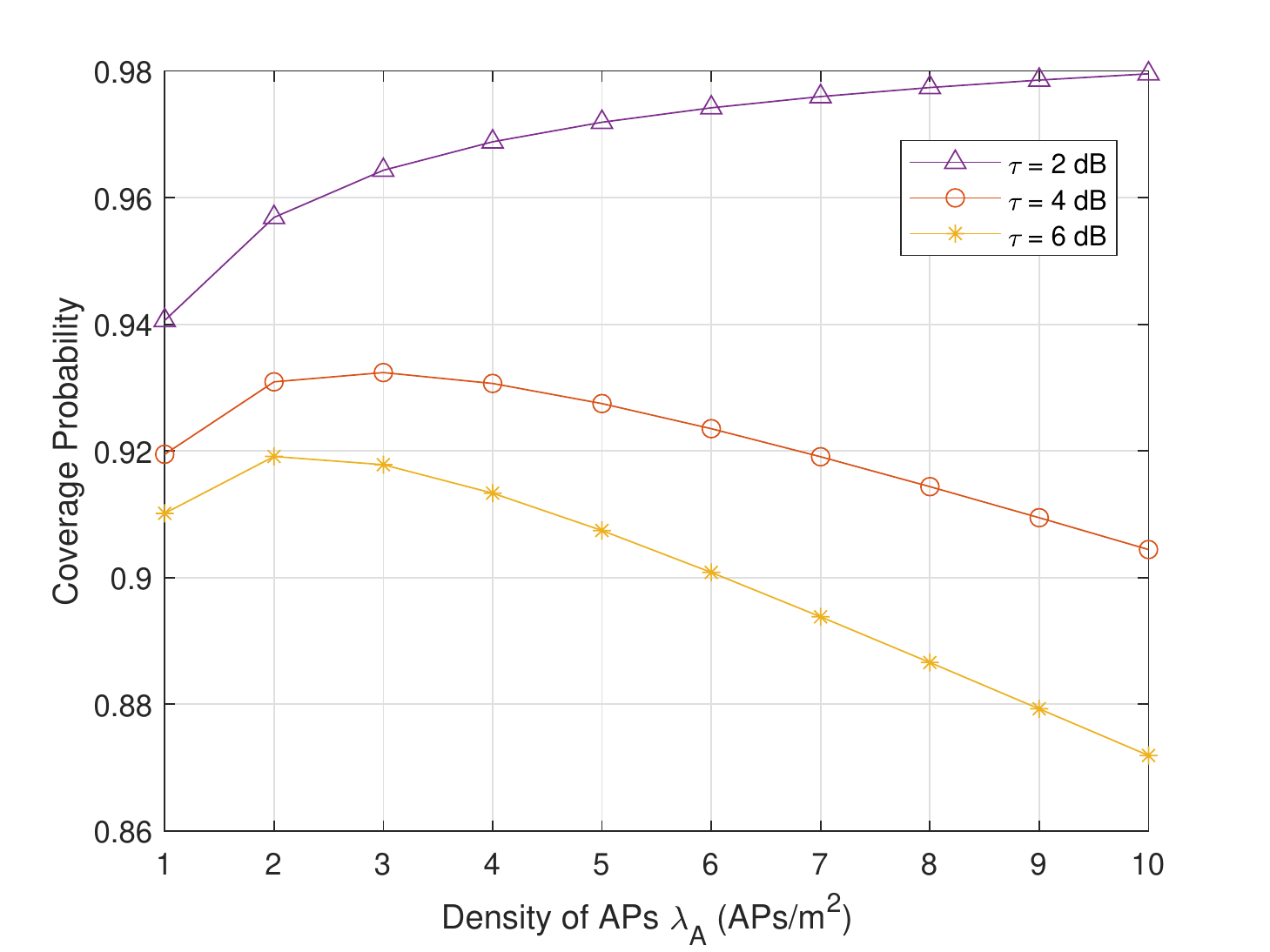}
    \caption{Coverage probability as a function of the density of APs and SIR threshold.}
    \label{fig3:b}
\end{figure}


\begin{figure}[t]
    \centering
    \includegraphics[scale=0.5]{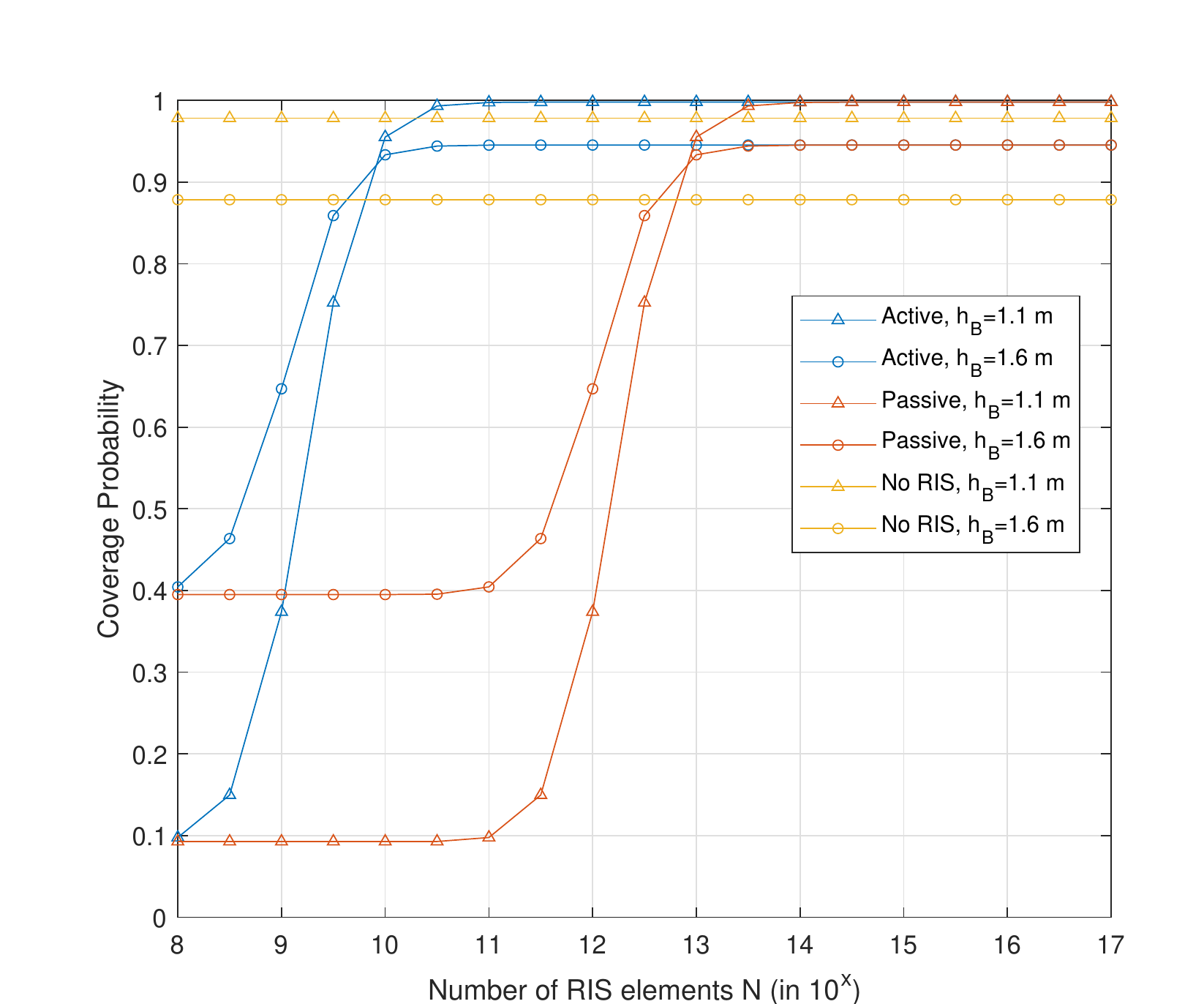}
    \caption{Coverage probability as a function of RIS elements and blockages height.}
    \label{fig3:c}
\end{figure}


\section{Numerical results}
\label{sec:results}
In this section, we validate the analytical results by conducting intensive Monte-Carlo simulations for an RIS-assisted indoor THz downlink network. Unless otherwise stated, we set $P_\text{A}=1$~mW, $N_\text{A}=10$~antennas, $f=0.3$~THz, $k_{a}(f)=0.075$~m$^{-1}$, $L_x$ = $L_y$ = $\frac{c}{2f}$,  $R_t=\sqrt{140}$~m, $h_\text{A}=3$~m, $h_\text{U}=1$~m, $h_\text{R}=0.75$ $h_\text{A}$, \textcolor{black}{3D location of RIS is $[1,1,h_{\text{R}}] \implies$} $v_{0}=\sqrt{2}$~m, $r_\text{B}=0.22$~m, $h_\text{B}=1.63$~m, $\lambda_{\text{B}}=2$~blockages/m$^2$, $\lambda_{\text{A}}=1$~APs/m$^2$, $G_\text{A}=G_\text{U}=30$~dB, $\tau=2$~dB, and $N=10^{13}$~RIS elements. 
We assume that every antenna is transmitting with unit energy i.e., $\vert f_j \vert^2=1, \forall j$. The Monte-Carlo simulations are conducted for $10^6$ realizations of the homogeneous PPP of APs and for $h_j$, $g_i$, $[\mathbf{H}]_{i,j} \sim C\mathcal{N}(0,1), \ \ \forall i \in \{ 1..N\}, \ j \in \{ 1..N_\text{A}\}$.

Fig.~\ref{fig:Ap_v_LB} and Fig.~\ref{fig:CP_v_LB} present the analytical (solid lines) and Monte-Carlo simulations (markers) results of the association and coverage probabilities as a function of the intensity of blockages $\lambda_\text{B}$. Fig.~\ref{fig:Ap_v_LB} and Fig.~\ref{fig:CP_v_LB} show a close match between the simulations and analytical results which validates the analysis developed in Section~\ref{sec:analysis}. As the density of blockages increases, the direct-only and RIS-only association probabilities start to increase then decrease gradually while association through composite links is always decreasing with $\lambda_{\text{B}}$. This is because a lower number of blockages provides a high chance of availability of both links (composite). However, with the existence of more blockages, the composite link association probability drops below the direct and RIS-only cases. As the RIS is deployed at a lower height than the APs, the RIS link is more likely to be blocked than the direct link. Thus, the direct association probability outperforms association through RIS. On the other hand, the overall coverage probability drops with the increase of $\lambda_{\text{B}}$ as more APs with good channel conditions will be blocked. Such behavior is different than the direct only case where the coverage probability shows an optimal $\lambda_{\text{B}}$ at which the coverage probability is maximized. 
\textcolor{black}{Moreover, to see the effectiveness of the association policy, we plot no association curve that shows the probability of the event that UE does not get any link to the nearest AP. One can see that even at $4$ average number of blockages in a meter-squared area the no-association probability is 15 $\%$, which is a reasonably low number. }


Fig.~\ref{fig3:a} shows the impact of the RIS height and distance with respect to the typical UE. We can clearly note that moving the RIS away from UE decreases the coverage probability due to the increased pathloss. 
Fig.~\ref{fig3:a} also shows that moving the RIS closer to APs provides better coverage probability for low UE-RIS distances. A higher RIS height increases the LoS probability of the RIS link and improves as a result the composite link coverage probability since an alternative link exists with a higher probability when the direct link is blocked. 

Fig.~\ref{fig3:b} highlights the impact of the density of APs on the overall coverage probability for different SIR thresholds. A higher density of APs increases the chance of having a nearby non-blocked AP to serve the UE, thus improving the coverage probability. This happens up to some extent after which increasing further the density of APs causes a degradation in the coverage probability due to the increased level of interference on both direct and RIS links. Thus, an optimal density of APs should be deployed to maximize the coverage probability. The optimal density of APs is larger for low SIR thresholds which show an improved performance compared to high SIR threshold values.  

Finally, Fig.~\ref{fig3:c} presents the coverage probability as a function of the number of RIS elements $N$ and the height of blockages and compare an RIS-assisted THz network with a passive RIS, an active RIS and the case of a THz network when no RISs are implemented. The active RIS is implemented by considering an extra gain $G_{\text{R}}=30$~dB provided by the existing RIS. As the number of RIS elements increases, the coverage probability increases till a certain point after which adding more RIS elements does not affect the overall coverage probability. { The reason behind this saturation is that the association probabilities are not functions of $N$, while the conditional coverage probabilities reach their maximum value of 1.} 
For a low number of RIS elements, a higher blockage height increases the coverage probability for the active and passive RIS cases. However, when $N$ increases, a higher blockages height negatively affects the coverage probability. { This is due to the fact that the composite link is significantly dominant for the blockages with lower height, while the direct link is dominant for the blockages with more heights.} 

Fig.~\ref{fig3:c} also shows that an active RIS starts providing coverage gains when the number of RIS elements exceeds $10^{10.5}$ for $h_{\text{B}}=1.1$~m and $10^{10}$ for $h_{\text{B}}=1.6$~m. On the other side, a passive RIS requires relatively larger numbers of RIS elements ($10^{13.5}$ for $h_{\text{B}}=1.1$~m and $10^{13}$ for $h_{\text{B}}=1.6$~m) to overcome the impact of increased pathloss and molecular absorption losses at THz frequencies. We note here that, although THz frequencies allow for adding more RIS elements with reasonable panel sizes, a large RIS panel is required to provide reasonable gains in THz networks. Furthermore, optimal RIS placement is required to mitigate the impact of blockages and increased losses at high frequencies.

\begin{figure}[t]
    \centering
    \includegraphics[scale=0.6]{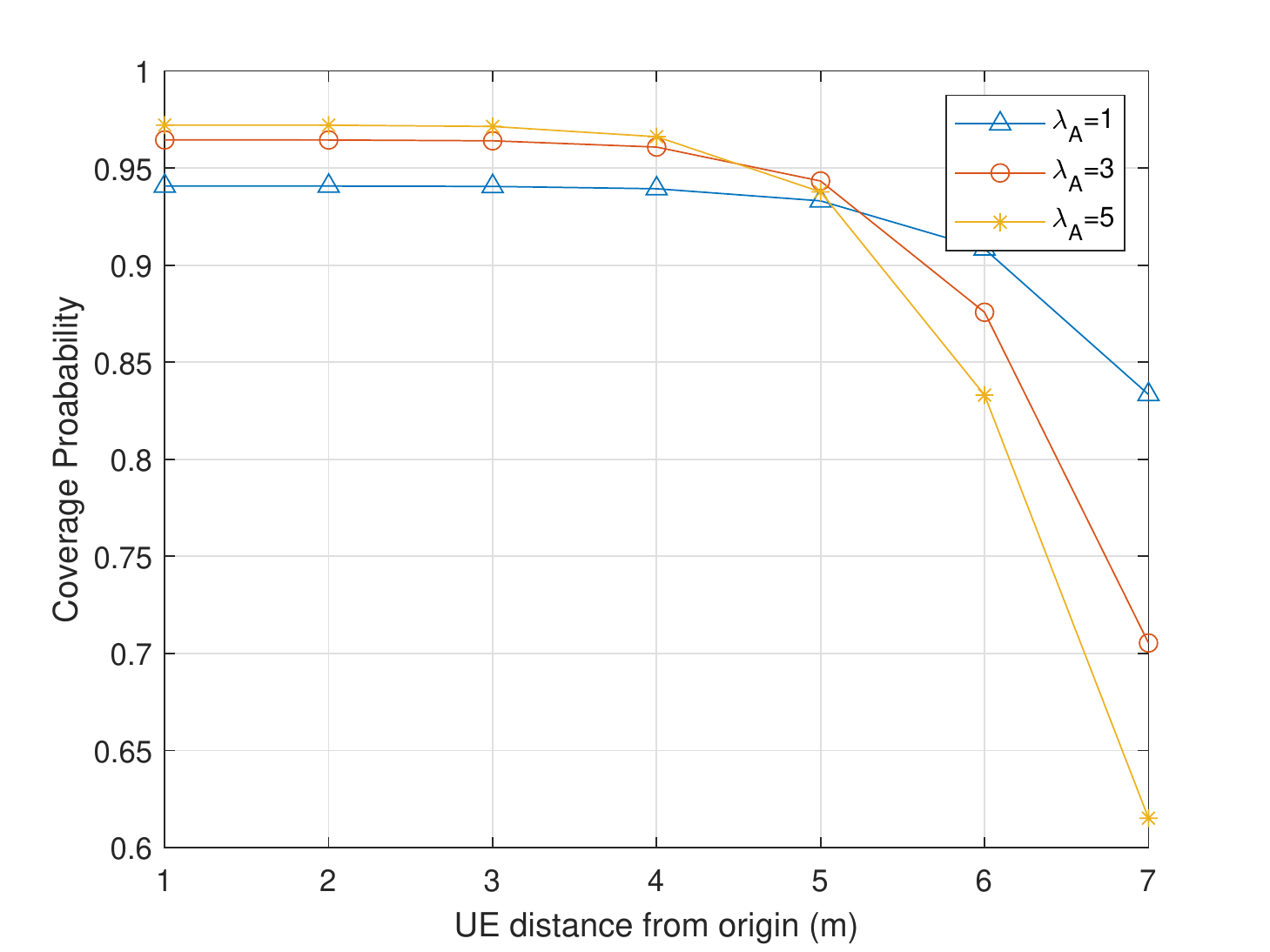}
    \caption{Coverage probability as a function of UE 2D distance.}
    \label{fig:PC_V_UE}
\end{figure}


\textcolor{black}{Fig.~\ref{fig:PC_V_UE} illustrates the impact of the user location on the coverage probability of the considered system. We can clearly see that, as the user is close to the origin, the coverage probability is almost constant. When the UE gets closer to the edge of the finite area, the coverage probability drops significantly. The degradation of the coverage probability at the edge of the network is caused by the reduced likelihood to find a close by AP to serve the user. Such behavior cannot be captured if the network is modeled as an infinite PPP, in which the interference seen by the user is the same across the network. One can also observe that the increase in the number of APs provides more interference at the origin and therefore a bit reduced coverage due to significant interference, while at the edge, more number of APs provides better coverage than less number of APs.}

\begin{figure}[t]
    \centering
    \includegraphics[scale=0.6]{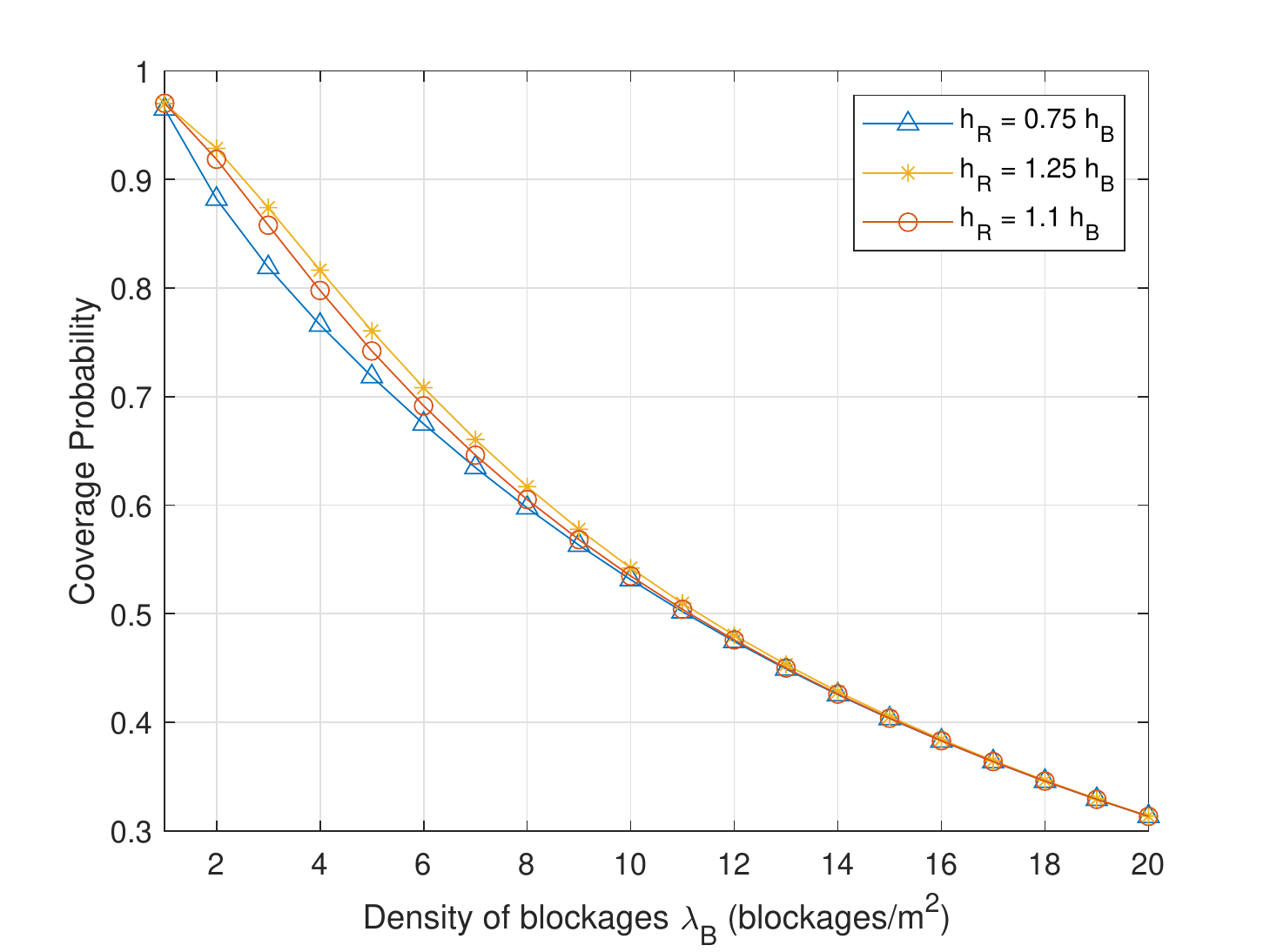}
    \caption{Coverage probability as a function of the density of blockages and RIS height.}
    \label{fig:Pcov_hR}
\end{figure}


{\color{black}Lastly, Fig.~\ref{fig:Pcov_hR} vividly illustrates the coverage probability as a function of the density of blockages and RIS height. Note that Fig.~\ref{fig:Pcov_hR} accounts for a specific scenario where the height of the RIS is relatively low, causing the blockages' height to surpass that of the RIS ($h_\text{R}=0.75$~ $h_{\text{B}}$). It is evident that under this condition, blockages exist on both the AP-RIS and RIS-UE links. Thus, the association probabilities with both the composite and RIS links significantly reduce, leading to their minimal contribution to the coverage probability. The direct impact of such behavior is an expected lower performance of the coverage probability compared to the higher RIS height cases, highlighting the critical impact of the efficient deployment of RIS to extend the coverage of THz networks.}
\vspace{-0.2cm}
\section{Conclusion}\label{sec:conclusion}
\vspace{-0.1cm}
We investigated the coverage performance of a RIS-assisted THz network with blockages. 
The obtained results highlight the importance of optimizing RIS deployment to extend the coverage of THz networks amid interference and blockages and identify clear trade-offs in terms of RIS sizes and energy consumption. 
\textcolor{black}{This work opens up many exciting directions for future work, e.g., optimal design of the RIS phase shift matrix $\mathbf{\Phi}$, incorporating realistic blockage models, studying the role of the RIS size on the LoS probability, assuming multiple antennas at the UE, studying other association policies, etc. }   





\vspace{-0.2cm}
\footnotesize{
\bibliographystyle{IEEEtran}
\bibliography{references}
}

\vfill\break

\end{document}